\documentclass[prl,twocolumn,superscriptaddress,amsmath,amssymb,aps,nofootinbib,10pt]{revtex4-2}

\usepackage[utf8]{inputenc}
\usepackage{amsmath,amsthm,amssymb,amsfonts,braket}
\usepackage{placeins}[verbose]
\usepackage{tabularx, longtable, multirow}
\usepackage{graphicx}
\usepackage{gensymb} 
\usepackage{dcolumn}
\usepackage{bm,dsfont}
\usepackage[english]{babel}
\usepackage{comment}
\usepackage[dvipsnames]{xcolor}
\usepackage{siunitx}
\usepackage[normalem]{ulem}
\usepackage[T1]{fontenc}
\usepackage{physics}
\usepackage{mathtools}

\def\bk{{\mathbf{k}}}

\def\bxi{{\boldsymbol{\xi}}}

\begin{document}

\title{Observation of a giant Goos-H\"anchen shift for matter waves}

\author{S. McKay}
\affiliation{Department of Physics, Indiana University, Bloomington IN 47405, USA}
\affiliation{Center for Exploration of Energy and Matter, Indiana University, Bloomington, 47408, USA}
\author{V.O. de Haan}
\affiliation{BonPhysics Research and Investigations BV, Laan van Heemstede 38, 3297AJ Puttershoek, The Netherlands}
\author{J. Leiner}
\affiliation{Neutron Sciences Directorate, Oak Ridge National Laboratory, Oak Ridge, TN, 37830, USA}
\author{S. R. Parnell}
\affiliation{Faculty of Applied Sciences, Delft University of Technology, Mekelweg 15, 2629 JB Delft, The Netherlands}
\affiliation{ISIS, Rutherford Appleton Laboratory, Chilton, Oxfordshire, OX11 0QX, UK}
\author{R. M. Dalgliesh}
\affiliation{ISIS, Rutherford Appleton Laboratory, Chilton, Oxfordshire, OX11 0QX, UK}
\author{P. Boeni}
\affiliation{Physics Department E21, Technical University of Munich, D-85748 Garching, Germany}
\affiliation{SwissNeutronics AG, Brühlstrasse 28, CH-5313 Klingnau, Switzerland}
\author{L.J. Bannenberg}
\affiliation{Faculty of Applied Sciences, Delft University of Technology, Mekelweg 15, 2629 JB Delft, The Netherlands}
\author{Q. Le Thien}
\affiliation{Department of Physics, Indiana University, Bloomington IN 47405, USA}
\author{D. V. Baxter}
\affiliation{Department of Physics, Indiana University, Bloomington IN 47405, USA}
\affiliation{Center for Exploration of Energy and Matter, Indiana University, Bloomington, 47408, USA}
\affiliation{Quantum Science and Engineering Center, Indiana University, Bloomington, IN 47408, USA}
\author{G. Ortiz}
\affiliation{Department of Physics, Indiana University, Bloomington IN 47405, USA}
\affiliation{Quantum Science and Engineering Center, Indiana University, Bloomington, IN 47408, USA}
\affiliation{Institute for Advanced Study, Princeton, NJ 08540, USA}
\affiliation{Institute for Quantum Computing, University of Waterloo, Waterloo, N2L 3G1, ON, Canada}
\author{R. Pynn}
\email{rpynn@iu.edu}
\affiliation{Department of Physics, Indiana University, Bloomington IN 47405, USA}
\affiliation{Center for Exploration of Energy and Matter, Indiana University, Bloomington, 47408, USA}
\affiliation{Neutron Sciences Directorate, Oak Ridge National Laboratory, Oak Ridge, TN, 37830, USA}
\affiliation{Quantum Science and Engineering Center, Indiana University, Bloomington, IN 47408, USA}

\date{\today}

\begin{abstract}
The Goos-Hänchen (GH) shift describes a phenomenon in which a specularly-reflected beam is laterally translated along the reflecting surface such that the incident and reflected rays no longer intersect at the surface.
Using a neutron spin-echo technique and a specially-designed magnetic multilayer mirror, we have measured the relative phase between the reflected up and down neutron spin states in total reflection. The relative GH shift calculated from this phase shows a strong resonant enhancement at a particular incident neutron wavevector, which is due to a waveguiding effect in one of the magnetic layers.
Calculations based on the observed phase difference between the neutron states indicate a propagation distance along the waveguide layer of \SI{0.65}{\milli \meter} for the spin-down state, which we identify with the magnitude of the giant GH shift.
The existence of a physical GH shift is confirmed by the observation of neutron absorption in the waveguide layer.
We propose ways in which our experimental method may be exploited for neutron quantum-enhanced sensing of thin magnetic layers.
\end{abstract}

\maketitle
\emph{Introduction.}---
The Goos-H\"anchen (GH) shift, in which incident and reflected rays do not exactly intersect
at the reflecting surface, was first experimentally confirmed for light in 1947 \cite{GH_1943,gh1}. Although the longitudinal GH shift for light beams is often of similar magnitude to the wavelength of the light used, much larger shifts as large as $\SI{1}{\milli \meter}$, dubbed \textit{giant GH shifts}, have recently been calculated for resonant structures \cite{Wu2019} and are now proposed as the basis for ultra-sensitive devices for measuring temperature \cite{Zhou2021} and relative humidity \cite{Wang2016}.

About a decade ago, de Haan \textit{et al.} \cite{neutronGH} attempted to confirm the existence of the GH effect for neutrons, the first type of matter-waves to be explored. Using the technique of neutron spin echo, they measured the difference in phase between the up and down neutron spin states reflected from a magnetic mirror made of Permalloy.
Unfortunately, the phase difference in this case was small, and critics suggested that the experimental result could have been caused by neutron depolarization \cite{Ignatovich_2010,dehaanResponse}. Furthermore, critics noted that the phase difference between states is not a direct confirmation of the relative GH spatial shift of the neutron spin states, which must be inferred from theory when only the relative phases are measured.
Theory predicted a maximum GH shift of $\SI{2.8}{\micro \meter}$ for the experiment reported in \cite{neutronGH}.
Several theoretical studies of the GH effect for neutrons have been published \cite{Ignatovich04,Frank14,Bushuev_Frank_2018}, at least one of which \cite{Frank14} proposed increasing the magnitude of the effect by using multilayered magnetic structures; we have adopted this approach using a specially-designed multilayer.

\emph{Sample design and experiment.}---
\begin{figure*}[t]
    \centering
    \includegraphics[width=\linewidth]{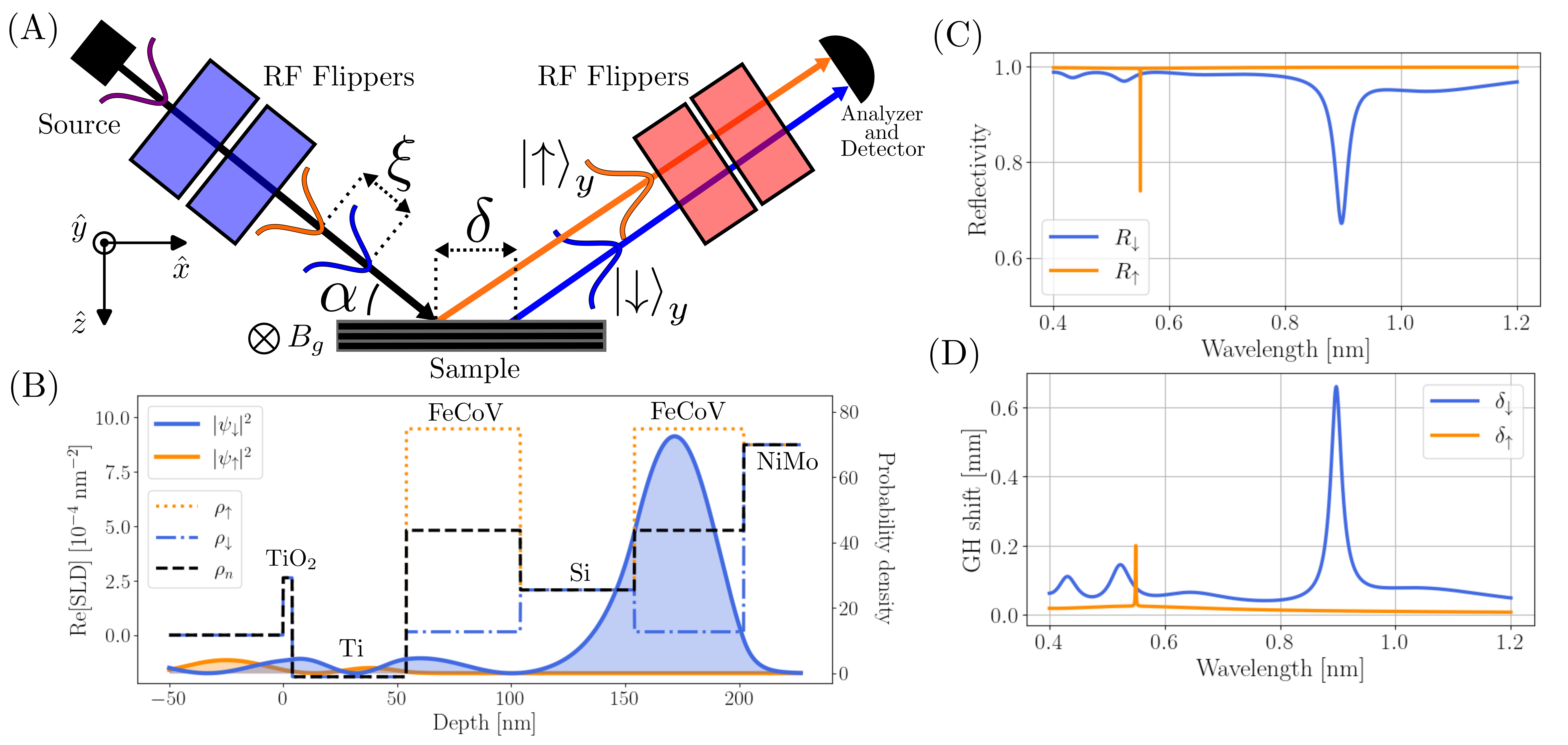}
    \caption{\label{fig:setup}
    (A) Sketch of the experimental setup. The neutron (purple) initially polarized in the $( \ket{\uparrow}_y + \ket{\downarrow}_y ) / \sqrt{2}$ state is spin-path entangled by the first set of radio-frequency (rf) flippers, resulting in an incident state with a wavepacket separation $\xi$. During reflection, the spin-up (orange) and spin-down (blue) state experience a different GH shift; the relative GH shift is labeled~$\delta$. Finally, the second pair of flippers disentangles the reflected state, which recombines the two wavepackets (producing a neutron spin echo when the reflecting sample is non-magnetic).
    The composition of the multilayer sample is displayed in Tab. \ref{tab:SLDs}. Experiments were performed with the magnetization of the sample in the $\hat{x}$ and $\hat{y}$ directions. The weak guide field at the sample position is labeled $B_g$. The incident grazing angle $\alpha$, GH shift, and wavepacket separation are all exaggerated for clarity.
    (B) Calculated probability densities of both spin states $\left| \psi_i \right|^2$ with $i \in \{ \uparrow,\downarrow \}$ (with $\hat{y}$ as the quantization axis) in the various layers relative to an incident planewave state with unit amplitude and $\lambda=\SI{0.9}{\nano \meter}$ using the Parratt formalism \cite{Parratt1954}. There is a strong resonance of the spin-down state in the lower FeCoV layer while the spin-up state has negligible probability density beyond the Ti layer. The spin-up, spin-down, and nuclear SLDs are labeled $\rho_{\uparrow}$, $\rho_{\downarrow}$, and $\rho_n$, respectively.
    (C) Calculated reflectivity $R_i$  for the up (orange) and down (blue) neutron spin states.
    (D) Calculated values of the GH shifts using the ACH formula.
    In parts (B-D), a grazing incidence angle of $\alpha = 0.34^{\circ}$ is used and absorption is included using the values of $\mathrm{Im}[\rho]$ displayed in Tab. \ref{tab:SLDs}. The correlation between the dip in reflectivity and the large GH shift seen in (C) and (D) is discussed in Ref.~\cite{SM}.}
\end{figure*}
As shown in Fig.~\ref{fig:setup} and Tab.~\ref{tab:SLDs}, our multilayer, fabricated by Swiss Neutrons AG, consists of the following nominal sequence, as seen by the neutrons incident from air: 50 nm Ti; 50 nm Vacoflux~50\textsuperscript{\textcopyright}; 50 nm Si; 50 nm Vacoflux~50; and 100 nm NiMo on a polished Si substrate (total size $100 \times 50 \times \SI{0.78}{\milli \meter}$).
The layer thicknesses were determined during fabrication and subsequently largely verified by X-ray reflectometry \cite{SM}. The capping Ti layer (which itself is topped with a few nm of native oxide presumed to be rutile TiO$_2$) is used to prevent oxidation of the uppermost Vacoflux layer. Vacoflux is a commercially-available FeCoV alloy (Vacuumschmelze GmbH \& Co. KG) with an elemental composition of 49\% Fe, 49\% Co, and 2\% V by weight and a bulk saturation magnetization of \SI{1.8e7}{\ampere \per \meter}.
Vacoflux has two important properties.
First, it is magnetically soft and, when deposited in a manner introducing appropriate strain, can be magnetized by a modest $\SI{20}{\milli \tesla}$ magnetic field along an in-plane easy axis, remaining fully magnetized when removed from the field \cite{Clemens1997}.
Second, the composition of Vacoflux leads to a neutron scattering length density (SLD) that is close to zero for spin-down neutrons whose magnetic moments are aligned with the magnetization direction in the alloy layer.
The NiMo layer contains $9.7~\pm~1.5\%$~Mo (atomic percent), which is sufficient to ensure that there is no magnetic neutron scattering from this layer \cite{Padiyath2004}. Because the Vacoflux layer closest to the substrate has positive SLD layers on either side (Si above and NiMo below), we anticipate a waveguiding, or Fabry-P\'{e}rot, effect for spin-down neutrons in this layer and no such effect for spin-up neutrons \cite{DeWames_Sinha_1973,Feng_1994}.

A sketch of the experimental geometry is shown in Fig.~\ref{fig:setup}(A). Our entangled-neutron reflection experiment used the Larmor beamline at the ISIS neutron and muon source in the United Kingdom. Two radio-frequency (rf) neutron spin flippers separated by a distance of $\sim \SI{1.5}{\meter}$ operating at a frequency of $\SI{1}{\mega \hertz}$ with the boundaries of their static magnetic fields perpendicular to the direction of neutron travel were used to mode-entangle (i.e., intra-particle entangle) the spin and path states of the incident neutrons \cite{contextuality,Kuhn2021,Leiner_2024}.
The two flippers before the sample induce a phase difference between the neutron spin states, which is reversed by two similar flippers after the scattering sample, thereby producing a neutron spin echo when the reflecting sample is non-magnetic \cite{Mezei_1972}.
This relative phase $\theta_k$, commonly called the \textit{Larmor phase}, is given by $\theta_k = \bm k \cdot \bm \xi   = 2 f m \lambda L / \hbar$, where the \textit{entanglement length} $\xi=|\bm \xi |$ is the wavelength-dependent spatial separation of the two entangled spin states after exiting the second rf flipper, $\bm k = (k_x,k_y,k_z)$ the neutron wavevector, $f$ the rf frequency, $m$ the mass of the neutron, $L$ the distance between each pair of rf flippers, and $\lambda$ the neutron wavelength~\cite{Golub_1987}.
For all measurements in this experiment, the spatial separation of the spin states was purely longitudinal (i.e., along the direction of travel $\hat{k}$ of the neutron), and the wavelength-dependent Larmor phase was fixed to $\theta_k~\approx~[\SI{4.8e4}{\nano \meter^{-1}}]~\lambda$, which also fixes $\xi~\approx~[\SI{7.6e3}{\nano \meter^{-1}}]~\lambda^2$.

\begin{table}[tb]
\centering
\newcolumntype{R}{>{\centering\arraybackslash}X}
\begin{tabularx}{.99\linewidth}{R|R|R|R|R}
Material & Thickness \newline {\footnotesize nm} & $\rho_n$ \newline {\footnotesize $10^{-4}$ nm$^{-2}$} & $\rho_m$ \newline {\footnotesize $10^{-4}$ nm$^{-2}$} & $-\mathrm{Im}[\rho]$ \newline {\footnotesize $10^{-7}$ nm$^{-2}$} \\ \hline
TiO$_2$  & 4 & 2.63 & 0 & 0.54 \\
Ti       & 50        & -1.91 & 0 & 0.96 \\
FeCoV    & 50        & 4.81  & 4.66 & 4.5 \\
Si       & 50        & 2.07  & 0 & 0.024 \\
FeCoV    & 50        & 4.81  & 4.66 & 4.5 \\
NiMo     & 100       & 8.73  & 0 & 1.0 \\
Si       & Substrate & 2.07  & 0 & 0.024 
\end{tabularx}
\caption{\label{tab:SLDs} 
Table of nominal layer thicknesses and nuclear ($\rho_n$), magnetic ($\rho_m$), and imaginary ($\mathrm{Im}[\rho]$) scattering length densities of the sample. The non-magnetic sample used for normalization of the echo polarization consisted of a \SI{100}{\nano \meter} thick NiMo layer on a Si substrate. The SLDs of the FeCoV layers were calculated assuming 95\% bulk density and 92\% of the bulk magnetic saturation (\SI{1.8e7}{\ampere \per \meter}).
The composition of the FeCoV (Vacuflux 50\textsuperscript{\textcopyright}) and NiMo alloys are given in the text.
As shown in Tab. \ref{tab:XRR} in \cite{SM}, the layer thicknesses were largely verified by X-ray reflectometry.
}
\end{table}

The pre-magnetized multilayer was mounted in reflection geometry with its magnetization $\bm{M}_{\mathrm{sam}}$ either parallel or perpendicular to a weak ($\sim~\SI{0.5}{\milli \tesla}$) magnetic guide field $\bm{B}_g = - B_g \hat{y}$ used to define the quantization axis for the neutron spins. This field had no effect on the sample magnetization. A grazing incidence angle of $\alpha=0.34 \pm .02^{\circ}$ was used during the experiment, ensuring that neutrons with $\lambda~\gtrsim~\SI{0.4}{\nano \meter}$ fell within the region of total external reflection of the multilayer.
A supermirror bender with its blades parallel to the neutron scattering plane of the experiment was used as a polarization analyzer and a linear, scintillation-based, position-sensitive, neutron detector with a 1D pixel size of \SI{0.64}{\milli \meter} in the neutron scattering plane was used.
The neutron beam dimension perpendicular to the scattering plane was \SI{20}{\milli \meter}. Two slits defined the width of the observed neutron beam in the $\hat{z}$ direction of Fig. \ref{fig:setup}(A); the first \SI{0.5}{\milli \meter} wide slit was positioned \SI{25}{\centi \meter} before the sample and the second \SI{2}{\milli \meter} wide slit \SI{35}{\centi \meter} behind the sample to limit background scattering.
The spin echo was established using a reflecting sample consisting of a $\SI{100}{\nano \meter}$ layer of NiMo on a silicon substrate; the NiMo composition and substrate dimension were identical to that used for the multilayer sample.

For values of the component $k_z=(2\pi/\lambda)\sin\alpha$ of the neutron wave vector perpendicular to the reflecting surface of a sample that are below the critical values for the spin states, neutrons are usually totally externally reflected.
For both measurements of our multilayer in the parallel $\left(\bm{M}_{\mathrm{sam}} \parallel \bm{B}_g \right)$ and perpendicular $\left(\bm{M}_{\mathrm{sam}} \perp \bm{B}_g\right)$ orientations, the incident neutron state is given by $\ket{\psi_{\mathrm{in}}} = ( e^{-i \theta_k /2} \ket{\uparrow}_y + e^{i \theta_k /2} \ket{\downarrow}_y) /\sqrt{2}$.
After reflection, the neutron state is determined by the application of the appropriate optical transfer matrix~$\mathbf{M}$:
\begin{align}
    \mathbf{M}^{\|} =& \, \mathrm{diag}\left( r_{\uparrow} e^{i \phi_{\uparrow}}, r_{\downarrow} e^{i \phi_{\downarrow}} \right), \quad \mathbf{M}^{\perp} = \, \mathcal{U}^{\dagger} \, \mathbf{M}^{\|} \, \mathcal{U},
\end{align}
where $\mathcal{U}$ is the appropriate change-of-basis matrix and $r_{\uparrow} e^{i \phi_{\uparrow}}$ and $r_{\downarrow} e^{i \phi_{\downarrow}}$ are the complex reflectances for the spin-up and -down states, respectively.
The reflectances for each spin state can be calculated using the standard Parratt formalism \cite{Parratt1954}, and the general framework of Ref.~\cite{Quan2024} was utilized to analyze measurements in the spin-echo modality \cite{SM}.

Assuming perfect beam polarization, when the magnetization of the sample is parallel to the magnetic guide field at the sample position, the experimentally-measured spin echo polarization $ P_z^{\|}$ and reflectance $R^{\|}$ are given in the plane-wave limit by
\begin{align} 
    P_z^{\|} =& \frac{2 r_{\uparrow} \, r_{\downarrow}}{r_{\uparrow}^2 + r_{\downarrow}^2} \cos(\phi_{\uparrow} - \phi_{\downarrow}), \quad R^{\|} = \frac{r_{\uparrow}^2 + r_{\downarrow}^2}{2}.
\end{align}
The Larmor phase $\theta_k$ varies extremely rapidly with neutron wavelength at an rf frequency of 1~MHz, and we lack the wavelength resolution ($\sim~\SI{5e-3}{\nano \meter}$) to resolve such a rapidly-oscillating phase \cite{SM}.
Therefore, the polarization and reflectance in the perpendicular case become
\begin{align}
    P_z^{\perp} =& \frac{2 r_{\uparrow} r_{\downarrow}}{\left( r_\uparrow + r_{\downarrow} \right)^2} \left[ 1 + \cos(\phi_{\uparrow} - \phi_{\downarrow}) \right], \quad R^{\perp} = R^{\|}.
\end{align}

Because the measured polarization contains the phase information from the spin states, we can extract the GH shift. According to the Artmann-Carter-Hora (ACH) theory of the GH shift, the lateral GH shift for each spin state is given by $\delta_i = (\partial \phi_i / \partial k_z)  \cot \alpha$, where $\phi_i$ for $i \in \{ \uparrow, \downarrow\}$ is the phase of the neutron \cite{artmann,carterhora}. As our experiment is only sensitive to the relative GH shift between the two spin states, we also define the relative GH shift as $\delta = \delta_{\downarrow} - \delta_{\uparrow}$, which is shown in Fig. \ref{fig:setup}(A).

\emph{Results.}---
\begin{figure*}[t]
    \centering
    \includegraphics[width=\linewidth]{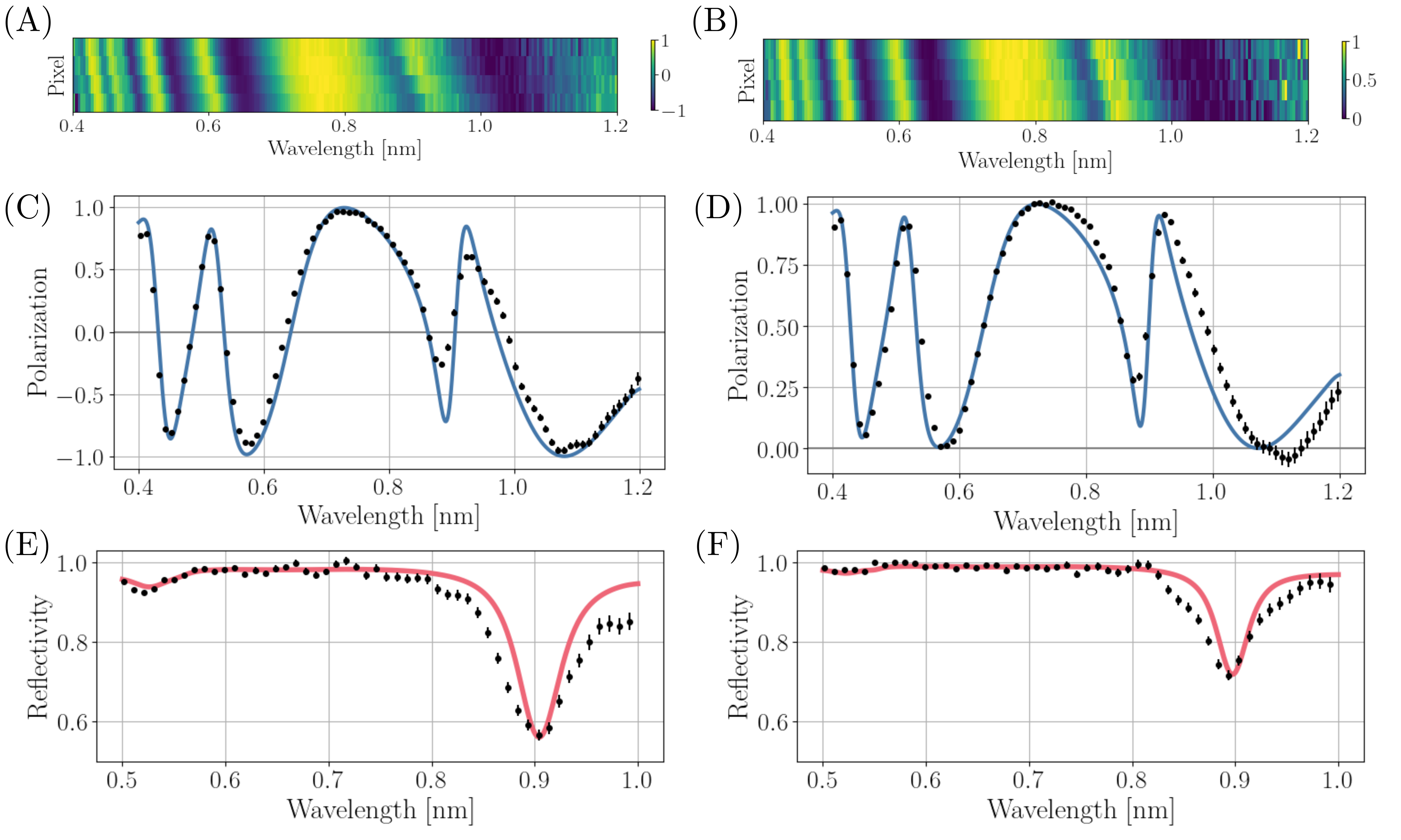}
    \caption{ \label{fig:data}
    (Top) Heat maps of the observed neutron polarization plotted against detector pixel and neutron wavelength:
    (A) $P_z^{\|}$, for the sample magnetization parallel to the guide field $\bm{B}_g$ [c.f. Fig. \ref{fig:setup}(A)], and (B) $P_z^{\perp}$, for the sample magnetization perpendicular to $\bm{B}_g$.
    (Middle) Measured polarization summed over the detector at constant momentum transfer: (C) $P_z^{\|}$  and (D) $P_z^{\perp}$. The blue curves in (C) and (D) represent the polarizations from Parratt simulations as described in the text.
    (Bottom) Measured sum of up- and down-spin state reflectivities: (E) $R^{\|}$ and (F) $R^{\perp}$. The red curves in (E) and (F) show the reflectivity results of the same simulations but with the imaginary part of the SLD in the second FeCoV layer increased by a factor of 7.5 in (E) and 2.5 in (F) from the table value of $\SI{-4.5E-7}{\nano \meter^{-2}}$.}
\end{figure*}
Even though the incident neutron beam in our experiment was well-collimated in the scattering plane, its angular divergence covered several detector pixels as shown in the false-color plot of echo polarization versus pixel and neutron wavelength $\lambda$ shown in Figs. \ref{fig:data}(A) and (B).
The fringes in Fig. \ref{fig:data}(A) and (B) are tilted because the measured neutron polarization depends on $k_z$.
Summing the measured neutron intensity at constant momentum transfer leads to the plots of echo polarization versus $\lambda$ shown in Figs.~\ref{fig:data}(C) and (D) for two orientations of the sample magnetization.

The curves in Figs.~\ref{fig:data}(C) and (D) represent a Parratt simulation convolved with the angular divergence of the neutron beam implied by the pixel size of the detector and the footprint of the neutron beam on the sample.
These simulations use the SLDs and layer thicknesses given in Tab.~\ref{tab:SLDs}, except for the case of the lower FeCoV layer, where we find a slightly better fit to the data with a thickness of 48 nm instead of the nominal 50 nm.
For the most part, the echo polarization is not particularly sensitive to the precise values of the SLDs given in Tab.~\ref{tab:SLDs}. The exception to this rule is the spin-down SLD of the lower FeCoV layer which determines the position of the sharp structure that occurs around \SI{0.9}{\nano \meter} in Figs.~\ref{fig:data}(E) and~(F). The value given in Tab.~\ref{tab:SLDs} is slightly less (92\%) than would be obtained using the bulk saturation magnetization of Vacoflux and has been chosen to give the correct position of the sharp jump in the echo polarization.
Between the two data sets, only $\alpha$ has been slightly adjusted by less than 0.01$^\circ$, within the potential error made in remounting the sample. Essentially, changing the incident angle translates the simulation curve right or left on the plot without changing its shape.
The data and the curves in Figs.~\ref{fig:data}(C) and (D) confirm that the phase difference between the two neutron spin states in the critical reflection region are indeed what one would expect from the plane-wave theory of neutron reflection.

Our experiment confirms that the $k_z$-dependence of the relative phase of the two neutron states is well-described by the usual optical theory. Thus we expect the GH shift produced by our sample to be accurately predicted by the ACH formula using values of the phases calculated from the Parratt simulation.
The GH shifts for up and down states calculated from the parameters of our multilayer are plotted for both spin states in Fig.~\ref{fig:setup}(D), showing a very large resonance for the down spin state at $k_z=\SI{0.04}{\nano \meter^{-1}}$ corresponding to $\lambda=\SI{0.9}{\nano \meter}$ at a grazing angle of incidence of $\alpha= 0.34^{\circ}$. At its peak, this resonance corresponds to a calculated GH shift of $\delta_{\downarrow}~\approx~\SI{0.65}{\milli \meter}$. 
The calculated GH shift for the spin-up state is negligible for $\lambda~\gtrsim~\SI{0.6}{\nano \meter}$, so we can approximate $\delta \approx \delta_{\downarrow}$.

While the measurement of the relative phases between the two states, taken together with the traditional ACH theory of the GH effect, predicts a giant GH shift for the down spin state, the experiment as described so far does not directly detect the shift. However, Vacoflux contains a substantial fraction (49\%) of cobalt which has a large neutron absorption cross section.
When this is included in the simulation, the phases of the two spin states change imperceptibly, but there is a significant change in the calculated reflectivity for the down spin state with no change for the up state as shown in Fig. \ref{fig:setup}(C).
Simulations verify that the lower FeCoV layer is the main contributor to the change in the down-state reflectivity. The reason that the lower FeCoV layer plays this important role is suggested by Fig. \ref{fig:setup}(B): the down-state SLD for FeCoV is close to zero and the lower FeCoV layer is surrounded on both sides by layers with positive SLDs. Thus, the lower FeCoV layer acts as a trapping potential for neutrons which are waveguided within this layer, parallel to the sample surface.
Because down-state neutrons see the lower FeCoV layer as a waveguide, they can be absorbed by the cobalt nuclei in this layer as they travel in the layer parallel to the sample surface. The resonant dip in the reflectivity of the down state corresponds to greatly increased absorption which, in turn, occurs because neutrons travel a large distance in the lower FeCoV layer.
The observation of the resonant dip in reflectivity in the critical region is thus direct evidence for a GH shift in our multilayer.

Figures~\ref{fig:data}(E) and (F) show plots of the sum of the measured up- and down-state reflectivities for our multilayer normalized to the average simulated values for wavelengths between 0.6~nm and 0.75~nm. The deep resonant dips corresponding to the giant GH shift are clear although the magnitudes of the dips depend on sample orientation.
To account for the observed depth of the dip using the Parratt simulation, we need to increase the imaginary part of the SLD in the lower FeCoV layer from the table value of $\SI{-4.5E-7}{}$ to $\SI{-3.4E-6}{\nano \meter^{-2}}$ when the sample magnetization is parallel to the magnetic guide field and to a value of $\SI{-1.1E-6}{\nano \meter^{-2}}$ in the perpendicular-magnetization case. Thus, while the dip in the reflectivity is a direct manifestation of the GH shift, its magnitude is not described by the standard optical theory incorporating neutron absorption alone.
However, as Sears \cite{Sears1982} has pointed out, the imaginary part of the SLD should include all channels which cause neutrons to be ``lost”, including both coherent and incoherent scattering in addition to absorption. Since the known incoherent scattering of FeCoV is small, we conclude that there is significant neutron scattering within the lower FeCoV layer and that this scattering depends on the direction in which the waveguided neutron travels in the layer.
This is hardly surprising given the manner in which the layer is intentionally strained during deposition in order to create a preferred direction for remanent magnetization. Significant small angle scattering, attributed to uncorrelated magnetic domains, has been reported for multilayers in which one component was an FeCoV layer of slightly different composition than that used in our experiment \cite{Kruijs2001}.

\emph{Conclusion.}---
In conclusion, we have provided strong experimental evidence for a giant Goos-H\"anchen shift for neutron matter waves.
The measured phase difference between up- and down-state reflected neutrons is well-described by usual optical theory (Parratt formalism), and the existence of a resonant dip in the reflectivity below the critical reflection wavevector confirms that down-state neutrons have an extended dwell time in our multilayer sample and thus a concomitant large GH shift \cite{SM}.
Based on the agreement of the measured relative phases and the standard optical theory we can use the ACH theory to estimate the GH shift as $\SI{0.65}{\milli \meter}$.

Finally, we note that measurements of the phase difference between up and down neutron spin states reflected from magnetic layered systems may be useful for some applications. For example, in entangled-beam reflectometry \cite{Quan2024} the phase difference can be measured for very thin and weakly magnetic layers below the critical angle where the reflectivity is unity for both neutron states. Standard spin asymmetry measurements used in polarized neutron reflectometry may be very hard to apply for such samples because the asymmetry occurs only at high momentum transfers where the reflectivity is very low. Thus, we conjecture that phase measurements below the critical edge may be exploited for enhanced quantum sensing of magnetism in two-dimensional systems.

\section*{Acknowledgements}
\begin{acknowledgments}

The IU Quantum Science and Engineering Center is supported by the Office of the IU Bloomington Vice Provost for Research through its Emerging Areas of Research program.  Some of this work was supported by the Department of Commerce through cooperative agreement number 70NANB15H259.
The experiment was performed on the Larmor instrument at the ISIS Neutron and Muon source (UK) supported by a beamtime allocation RB2310496 from the Science and Technology Facilities Council \cite{UK_beamtime}.
We would like to thank Amy Navarathna from the TU Delft for performing the X-ray reflectometry measurements.
This research was undertaken thanks in part to funding from the Canada First Research Excellence Fund. G.O. gratefully acknowledges support from the Institute for Advanced Study. Q.L.T. would like to thank Dennis Krause for pointing out the connection of this work with interferometry of unstable particles.

\end{acknowledgments}

\FloatBarrier
\bibliographystyle{apsrev4-2.bst}
\bibliography{sources.bib}

\clearpage
\newpage
\section*{Supplemental material}

\subsection{Layer thicknesses verification via X-ray reflectometry}

\begin{figure}[ht]
    \centering
    \includegraphics[width=0.99\linewidth]{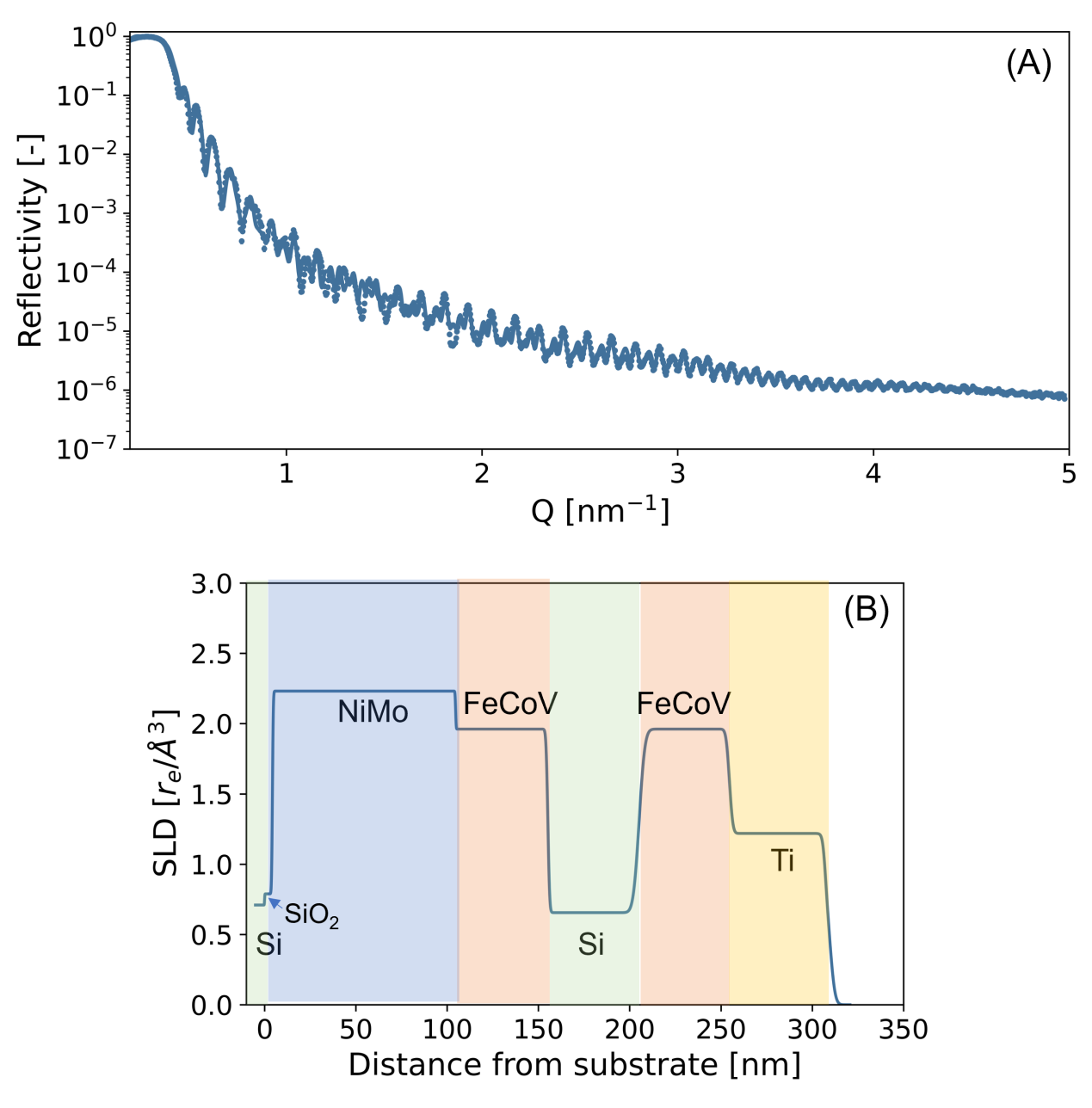}
    \caption{ \label{fig:XRR} (A) X-ray reflectometry data. The points indicate the measured data and the continuous line the fit to the data. (B) The corresponding scattering length density (SLD) profile.
}
\end{figure}

X-ray reflectometry (XRR) measurements were performed with a Bruker D8 Discover diffractometer equipped with a Cu K$\alpha$ source ($\lambda$~=~0.1542~nm, 40~kV and 40~mA). The X-ray diffractometer was operated with a parallel beam mirror and a 0.1 mm exit slit on the primary side, and two 0.1 mm fixed slits on the secondary side.
The LYNXEYE-XE detector was operated in high count rate 0D mode. To avoid saturation of the detector, two measurements were performed: one from 0~$\leq$~2$\theta$~$\leq$~2\degree~with a 0.1~mm Cu attenuator and from 0.8~$\leq$~2$\theta$~$\leq$ 7\degree~ in steps of 0.002\degree~ without an attenuator. These measurements were stitched using an in-house developed Python code. Data fitting was performed with GenX~3.6.20~\cite{glavic2022genx} and the default `Log' minimization function (figure of merit, FOM) was used. 

Figure \ref{fig:XRR} presents the XRR data, the corresponding fits and the scattering length density (SLD) profile. The fitted parameters are displayed in Table \ref{tab:XRR}.
In the fits, the theoretical sample length, beam size and resolution of the machine were provided as input and the thickness and roughness of all layers except the Si layer were fitted. The Si layer thickness was kept to the expected value of 50 nm. When fitting the Si layer thickness, a better fit could be obtained using a thickness of 42~nm. However, a silicon thickness of 42 nm is completely incompatible with our neutron data which is well described by a thickness of 50 nm. Since the growth rates of the layers were calibrated prior to deposition of the multilayer, we suspect that the silicon thickness deduced from fitting the X-ray result may be compensating for other small variations in the multilayer parameters that were not fitted.

With the exception of FeCoV, all densities were kept at the literature value. Fitting these densities did not result in a significant improvement of the fit nor a large change in the fitted densities. The FeCoV density was fitted, yielding a 6\% smaller value than the reported bulk density. When the FeCoV density was constrained to the bulk density, the fit became significantly worse. The XRR data were relatively insensitive to the density of the silicon layer. 

\begin{table}[h]
\centering
\newcolumntype{R}{>{\centering\arraybackslash}X}
\begin{tabularx}{.99\linewidth}{R|R|R|R}
Material & Thickness (nm) & Density \newline (F.U. /nm$^{3}$) & Roughness \newline (nm) \\ \hline
TiO$_2$  & 2.4 $\pm$ 0.4 & 31 & 2.3 \\
Ti       & 52.5 $\pm$ 0.2        & 56 & 1.0 \\
FeCoV    & 50.5 $\pm$ 0.2       & 80 & 0.8 \\ 
Si       & 50 (fixed)       & 50 & 2.4 \\
FeCoV    & 49.7 $\pm$ 0.3       & 80 & 1.5 \\
NiMo     & 100.3 $\pm$ 0.3      & 85 & 0.3 \\
SiO$_2$     & 4.3 $\pm$ 0.4      & 26 & 0.4 \\
Si       & Substrate & 50  & 0.1
\end{tabularx}
\caption{\label{tab:XRR} 
Table of fitted thicknesses, densities (expressed as number of formula units per nm$^3$), and roughness of each of the layers. The corresponding fits are shown in Fig.~\ref{fig:XRR}.}
\end{table}

\subsection{Neutron reflectometry data treatment}

We now discuss the specific data treatment used to generate the plots in Fig. \ref{fig:data} in the main text.
When using a time-of-flight neutron source, the momentum transfer~$q$ is a function of the spatial coordinate of the detector pixel $z$ and neutron wavelength $\lambda$; to first order,
\begin{equation} \label{eq:q conversion}
    q(z,\lambda) = \frac{4 \pi}{\lambda} \left[ \alpha - \frac{(z-z_0)}{2 L_d} \right],
\end{equation}
where $\alpha$ is the grazing angle of incidence (see Fig. \ref{fig:setup} in the main text), $z_0$ the center of the specularly reflected beam at the detector, and $L_d = \SI{4.35}{\meter}$ the distance from sample to detector. The intensities for both the up and down spin states were binned at constant $q$ independently.

After binning at constant $q$, each pixel of the NiMo sample polarization $P_0$ was well-fit by a second-degree Chebyshev polynomial of the first kind; this fit was used as the normalization in Figs.~\ref{fig:data}(C) and (D) in the main text.
The corresponding total intensity $I_z$ for each pixel of the NiMo reference sample was well-fit by the function $\log~I_z~=~a~\lambda~+~b$ in the region $\lambda \in [0.5,1.0] \, \SI{}{\nano \meter}$. 
The purpose of this fit was to remove the effect of the incident neutron spectrum; the overall scale of the data was set by the average value of the total intensity from the Parratt simulation in the region $\lambda \in [0.6,0.75] \, \SI{}{\nano \meter}$, which was chosen to be sufficiently far from resonance.
The data shown in Figs.~\ref{fig:data}(E) and (F) in the main text use this fit.

\begin{figure}[ht]
    \centering
    \includegraphics[width=\linewidth]{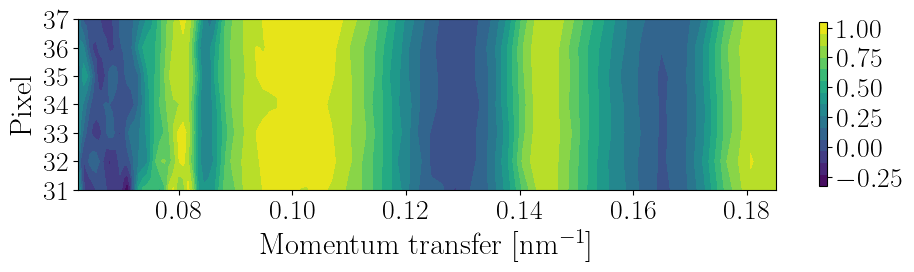}
    \includegraphics[width=\linewidth]{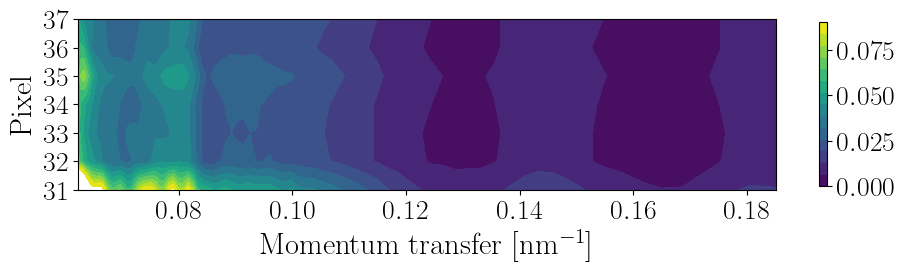}
    \caption{ \label{fig:2d pol perp}
    Normalized measured polarization $P_z^{\perp} / P_0$ (top) and error (bottom) as a function of pixel and momentum transfer.
    Pixel 34 corresponds to the center of the detector [$z_0$ in Eq.~\eqref{eq:q conversion}]. The pixel size is 0.64 mm.}
\end{figure}

\begin{figure}[ht]
    \centering
    \includegraphics[width=\linewidth]{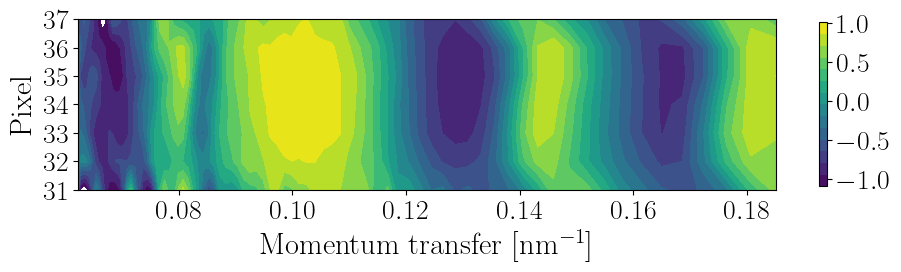}
    \includegraphics[width=\linewidth]{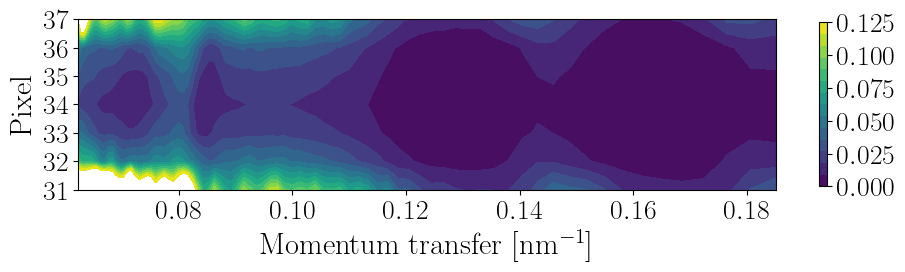}
    \caption{ \label{fig:2d pol para}
    Normalized measured polarization $P_z^{\|} / P_0$ (top) and error (bottom) as a function of pixel and momentum transfer.
    Pixel 34 corresponds to the center of the detector [$z_0$ in Eq.~\eqref{eq:q conversion}]. The pixel size is 0.64 mm.}
\end{figure}

The normalized spin echo polarization vs pixel and momentum transfer for the measurements where the sample magnetization was perpendicular and parallel to guide field are shown in Figs. \ref{fig:2d pol perp} and \ref{fig:2d pol para}, respectively.
The purpose of both the polarization and intensity fitting was to reduce the spurious noise that arises from point-by-point division of multilayer sample data by the NiMo reference sample data.

\subsection{Dwell Time Discussion}

A comparison of Figs.~\ref{fig:setup}(C) and (D) in the main text indicate that the simulated reflectivity profile of the down spin state closely mimics an inverted version of calculated GH shift for this state. To understand this, we note that the specular neutron reflectivity $R$ of a medium is related to the neutron absorption $A$ by the equation $R = 1-A$ in the critical-reflection region.
For a bulk material, it is easy to verify using the Fresnel form for the reflectivity and the ACH formula that to lowest order, $A \propto {\rm Im}[\rho] \delta$ where ${\rm Im}[\rho]$ denotes the imaginary part of the SLD. However, we have not been able to find any such simple relationship for a general multilayer; the Parratt simulation of our multilayer indicates that such a linear dependence of $A$ on ${\rm Im}[\rho]$ is only valid for values of ${\rm Im}[\rho]$ that are smaller than that implied by the neutron absorption in FeCoV.
One way to see that we might expect the absorption to be related to the GH shift for a multilayer follows from an equation for the absorption described in more detail by Piscitelli \textit{et al.} \cite{Piscitelli2015}:
\begin{eqnarray} \label{absorption}
    A(k_z)=-\frac{4\pi}{k_z} \int dz \, \frac{|\psi(z)|^2}{|\psi_i|^2} {\rm Im}[\rho(z)],
\end{eqnarray}
where $\psi(z)$ is the neutron wavefunction in the multilayer, $\psi_i$ is the incident neutron wavefunction, and the integration extends over the thickness of the multilayer. The review by Hauge and St{\o}vneng \cite{Hauge1989} discusses many past attempts to formulate expressions for the interaction time of a quantum particle with a one-dimensional potential. One possible expression for the \textit{dwell time} $\tau$ discussed by these authors due to B{\"u}ttiker \cite{Büttiker_1983} is
\begin{eqnarray} \label{dwelltime}
    \tau=\frac{1}{|\psi_i|^2 \, v_z}  \int dz \, |\psi(z)|^2,
\end{eqnarray}
where $v_z$ is the $z$-component of the incident neutron group velocity, and the integration bounds extend over the entire multilayer.
Physically, the dwell time is a measure of the time that the neutron spends in the multilayer, averaged over all incoming neutrons and scattering channels.
As shown in Fig.~\ref{fig:setup}(B), the squared wave function in our multilayer displays a very large peak in the lower FeCoV layer, so most of the integral in Eqs. \eqref{absorption} and \eqref{dwelltime} is concentrated in this region. Since the imaginary part of the SLD in that layer is constant, we can remove ${\rm Im}[\rho(z)]$ from the integral in Eq. \eqref{absorption} and combine Eqs. \eqref{absorption} and \eqref{dwelltime}.
Assuming that the GH shift can be written as $\delta=v_x \tau$, where $v_x$ is the group velocity of the neutron parallel to the reflecting surface, we obtain
\begin{eqnarray} \label{shift} 
    \delta = -\frac{A}{2\lambda {\rm Im}[\rho(z)]}.
\end{eqnarray}

Equation \eqref{shift} provides an intuitive justification for the relationship between the GH shift and absorption and implies that the effect of a GH shift will be seen in a departure of the sample's reflectivity from unity below the critical value for $k_z$.
Parratt simulations of our multilayer disagree quantitatively with Eq. \eqref{shift} by about 30\% even at small values of ${\rm Im}[\rho(z)]$, so the Eq. \eqref{shift} cannot be used to deduce an accurate value of the GH shift from the measured absorption. These simulations also show that the absorption at the position of the largest GH shift ($\lambda = \SI{0.9}{\nano \meter}$ in our experiment) can be described by the equation
\begin{equation}
    A = 1 - \exp \left( \left[\SI{-8.9e5}{\nano \meter^2}\right] \mathrm{Im}[\rho] \right).
\end{equation}
Since the neutron is traveling in the lower FeCoV waveguiding layer, we may expect that $A=1-\exp(-N \sigma_a \ell)$, where $N$ is the number density of absorbing nuclei in the layer, $\sigma_a$ is their effective absorption cross section, and $\ell$ is the distance travelled in the layer.
Since $\mathrm{Im}[\rho]=N\sigma_a/(2\lambda)$~\cite{Sears1982}, we can immediately see that $\ell=\SI{0.5}{\milli \meter}$ at $\lambda=\SI{0.9}{\nano \meter}$, which is very close to the value of the GH shift obtained from the ACH formula of \SI{0.65}{\milli \meter}.

\subsection{Theoretical methods}

We discuss next how the polarization and reflectivity for layered samples are calculated when the spin-echo technique is employed.
First, we take the incoming spin state to be $\ket{\uparrow}_z$. The neutron is then subject to a pair of rf flippers which act as an \textit{entangler} in the spin and momentum subsystems, followed by reflection off the sample, and finally is acted upon by another pair of rf flippers which act as a spin-momentum \textit{disentangler}.
The entangler can be represented by the operator \cite{Quan2024}
\begin{eqnarray}\hspace*{-0.6cm}
    \mathcal{U}^{\sf se}_{\hat{{\sf e}}}(\bxi) &=& \!\!\int \!d \bk  \ketbra{\bk}{\bk} \otimes \left(e^{-i \theta_k/2} |\uparrow\rangle_{\!{\hat{\sf e}}} \!\langle \uparrow|+  e^{i \theta_k/2}|\downarrow\rangle_{\!{\hat{\sf e}}} \!\langle \downarrow| \right),  \label{spin echo momentum}
\end{eqnarray}
where $\hat{\sf e}$ ($\hat{y}$ in our setup) is the entangler spin basis and $\theta_k =\bk\cdot\bxi$, where $\bxi$ represents the spin-echo length vector; in our experiment $\bxi$ is longitudinal. Equivalently, 
\begin{equation}
    \mathcal{U}^{\sf se}_{\hat{{\sf e}}}(\bxi) = \!\!\int \!d \bk  \ketbra{\bk}{\bk} \otimes \left(\begin{array}{cc}
        \cos \theta_k & \sin \theta_k  \\
        -\sin\theta_k & \cos \theta_k
    \end{array} \right).
\end{equation}
The disentagler is represented by $ \mathcal{U}^{\sf se}_{\hat{{\sf e}}}(\bxi)^\dagger$. 

The reflection operator $\mathbf{M}$ is computed, for the given sample multi-layer structure, via the spin transfer matrix formalism \cite{Blundell92,Pleshanov96,Quan2024}. When the sample magnetization direction is $\hat n =\hat{y}$ (i.e., the parallel case),
\begin{equation}
    \mathbf{M}^\| = \left( \begin{array}{cc}
         r_\uparrow e^{i\phi_\uparrow} + r_\downarrow e^{i\phi_\downarrow} & i( r_\uparrow e^{i\phi_\uparrow} - r_\downarrow e^{i\phi_\downarrow})  \\
         - i( r_\uparrow e^{i\phi_\uparrow} - r_\downarrow e^{i\phi_\downarrow}) & r_\uparrow e^{i\phi_\uparrow} + r_\downarrow e^{i\phi_\downarrow} 
    \end{array} \right),
\end{equation}
where $r_{\uparrow\downarrow}$, and $\phi_{\uparrow\downarrow}$ are $\bk$-dependent.
When $\hat n = \hat{x}$ (i.e., the perpendicular case),
\begin{equation}
    \mathbf{M}^\perp = \left( \begin{array}{cc}
         r_\uparrow e^{i\phi_\uparrow} + r_\downarrow e^{i\phi_\downarrow} &  r_\uparrow e^{i\phi_\uparrow} - r_\downarrow e^{i\phi_\downarrow}  \\
         r_\uparrow e^{i\phi_\uparrow} - r_\downarrow e^{i\phi_\downarrow} & r_\uparrow e^{i\phi_\uparrow} + r_\downarrow e^{i\phi_\downarrow} 
    \end{array} \right).
\end{equation}
We make the further assumption that the reflection operator does not vary significantly over the beam wavevector bandwidth~$\Delta k/k \sim$ 1\%. 

Before obtaining the polarization and reflectivity, we first note that since the effective optical potential in the FeCoV layer contains an imaginary part due to a non-zero neutron absorption cross-section and diffuse scattering, the reflection process is inherently non-unitary.
Therefore, we adopt the description of non-unitary quantum mechanics used widely in describing interferometry with unstable particles \cite{KRAUSE,karamitros}. The observed polarization is conditional upon the fact that the neutron has not been absorbed. For a perfectly collimated beam, this necessitates a normalization factor to the un-normalized density matrix $\rho(\bk')=\langle \bk' \ketbra{\Psi^{\sf se}}\bk'\rangle$:
\begin{equation}
    \rho(\bk') \rightarrow    \tilde\rho(\bk')=\frac{\rho(\bk')}{\Tr\rho(\bk')} , \ \Tr \Tilde\rho(\bk')=1 ,
\end{equation}
where the trace, $\Tr$, is performed over the spin-1/2 basis and
\begin{equation}
    \ket{\Psi^{\sf se}}=  \mathcal{U}^{\sf se}_{\hat{{\sf e }}}(\bxi)^ \dagger \, \mathbf{M}^{\|/\perp} \, \mathcal{U}^{\sf se}_{\hat{{\sf e}}}(\bxi) \ket{\uparrow}_z \otimes \ket{\bk'}
\end{equation}
is the un-normalized neutron spin-echo state after passing through the disentangler. Furthermore, the neutron beam has an  (incoherent) distribution $p(\bk')$ of momenta centered about $\bk$ with a width $\Delta k$ of order $\Delta k/k \sim$  1\%. Hence, the normalized density matrix becomes
\begin{equation}
    \hat \rho (\bk) =  \int d\bk' \ p(\bk') \tilde \rho(\bk').
\end{equation}

When the entangler and disentangler quantization axes ($\hat{\sf e}=\hat{y}$) are aligned with the sample magnetization ($\hat n = \hat y$), the measured polarization and reflectivity are $\bxi$-independent since $[\mathbf{M}^\|,\mathcal{U}^{\sf se}_{\hat{{\sf e}}}(\bxi)]=0$. Then, 
the polarization $P^\|_\mu=\Tr[\hat{\rho}(\bk) \hat{\sigma}_\mu]$ with $\mu \in \{x,y,z\}$ is given by
\begin{eqnarray} {\hspace*{-0.7cm}}
    \label{polarization simplified2}
    P^\|_x (\bk) &=& \frac{2 r_\uparrow (\bk)r_\downarrow(\bk) }{ r_\uparrow^2 (\bk ) + r_\downarrow^2 (\bk )}  \sin (\phi_\uparrow(\bk ) - \phi_\downarrow(\bk) ), \nonumber  \\
    P^\|_y(\bk) &=&  \frac{r_\uparrow^2 (\bk) - r_\downarrow^2 (\bk )}{ r_\uparrow^2 (\bk ) + r_\downarrow^2 (\bk )},\\
    P^\|_z(\bk ) &=&  \frac{2r_\uparrow (\bk )r_\downarrow (\bk )}{ r_\uparrow^2(\bk) + r_\downarrow^2(\bk)}  \cos (\phi_\uparrow (\bk ) - \phi_\downarrow (\bk ) )  \nonumber, \\
    R^\| (\bk) &=&  \int d\bk \,p(\bk)\Tr  \rho(\bk) = \frac{ r_\uparrow^2 (\bk ) + r_\downarrow^2 (\bk) }{2}.\nonumber
\end{eqnarray}
In this case, the distribution $p(\bk')$ acts effectively as a dirac delta function $\delta^3(\bk'-\bk)$ since $\Tilde {\rho}(\bk)$ is $\bxi$-independent. Therefore, the observed polarization $P_{\mu}^\|(\bk)$ and reflectivity $R^\|(\bk)$ do not differ from those originating from a perfectly collimated beam with wave vector~$\bk$.  

\begin{figure}[h]
    \includegraphics[width=\linewidth]{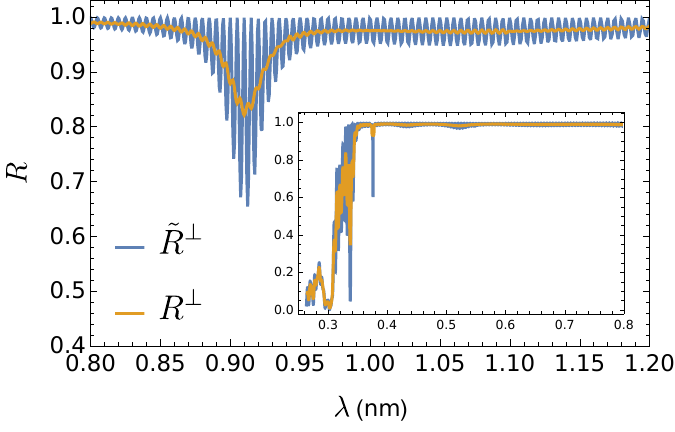}
    \caption{ \label{fig:spin-echo length oscillation} Theoretical prediction using Eq.~\eqref{polarization spin echoed} with the normalization $\Tr\rho$ and Eq.~\eqref{polarization spin echoed average} of sample's reflectivity in the perpendicular set up ($\hat{n}=\hat{x}$ and $\hat{\sf e}=\hat{y}$) with $\xi = \SI{200}{\nano \meter}^{-1}\lambda^2$ for a perfectly collimated beam ($\tilde R^\perp$) and a beam with $\Delta k / k \sim 1\%$ ($R^\perp$). We note that this value of $\xi$ is less than 10\% of the one used in our experiment; this small value of $\xi$ was chosen to highlight the oscillations.}
\end{figure}

In contrast, when the entangler and disentangler quantization axes are not aligned with the sample magnetization (i.e., ${\hat{\sf e}}\ne \hat n$), matters become mathematically cumbersome because $[\mathbf{M}^\perp,\mathcal{U}^{\sf se}_{\hat{{\sf e}}}(\bxi)]\neq0$ \cite{Quan2024}.
As an example, consider the case $\hat{\sf e}=\hat y$ and $\hat n=\hat x$. Then, the un-normalized polarization $\tilde P^\perp_\mu(\bk) = \Tr[\rho (\bk) \hat{\sigma}_\mu]$ and reflectivity $\tilde R^\perp (\bk) = \Tr \rho(\bk)$, for a perfectly collimated beam with the wave vector $\bk$, is given by  
\begin{widetext}
\begin{eqnarray}
   \tilde P^\perp_x(\bk) &=& \frac{1}{2} \left( \cos \theta_k \left(\sin \theta_k \left(r_\uparrow^2 (\bk) + r_\downarrow^2(\bk) -2r_\downarrow (\bk) r_\uparrow(\bk) \cos (\phi_\uparrow (\bk) - \phi_\downarrow (\bk) ) \right)   +r_\uparrow^2 (\bk) - r_\downarrow^2 (\bk) \right) \right) , \nonumber\\
   \tilde P^\perp_y(\bk)&=& r_\downarrow (\bk) r_\uparrow (\bk) \cos \theta_k \sin (\phi_\uparrow (\bk) - \phi_\downarrow(\bk) )  , \\
   \tilde P^\perp_z(\bk) &=& \frac{1}{2} \sin \theta_k \left(\left( r_\uparrow^2 (\bk) + r_\downarrow^2 (\bk)\right) \sin \theta_k +r_\uparrow^2 (\bk) -r_\downarrow^2(\bk)\right) + r_\downarrow (\bk) r_\uparrow (\bk) \cos ^2\theta_k \cos (\phi_\downarrow(\bk) -\phi_\uparrow (\bk))  ,    \nonumber \\
   \tilde R^\perp(\bk)&=&   \frac{1}{2 } \left( r_\uparrow^2 (\bk) + r_\downarrow^2 (\bk) + ( r_\uparrow^2 (\bk) - r_\downarrow^2 (\bk) ) \sin \theta_k \right) \nonumber.
    \label{polarization spin echoed}
\end{eqnarray}     
\end{widetext}
One can obtain the normalized polarization by dividing by a factor of $\Tr \rho(\bk)$.
We see that different from the case where the sample's magnetization is aligned with ${\hat{\sf e}}$, the $\theta_k$-dependence indicates that even for a perfectly collimated beam, both polarization and reflectivity display modulations with the spin-echo length $\xi$.
As illustrated in Fig.~\ref{fig:spin-echo length oscillation}, the observed reflectivity $R^\perp(\bk)$ for a beam with finite spread $p(\bk')$ has an oscillation with period $2\pi/\xi$ convolved with the reflectivity $\tilde{R}^\perp(\bk)$ of a perfectly collimated beam. Note that this average over $p(\bk')$ changes the observed polarization and reflectivity most significantly when their oscillation amplitudes vary the most over the oscillation period.

For our beam, the spread of $p(\bk')$ with $\Delta k/k \sim 1\%$ is $\sim \SI{1e8}{\meter^{-1}}$, while the oscillation characteristic length scale in $\theta_k$ is $2\pi/\xi \sim \SI{1e6}{\meter^{-1}}$; this is indeed $\sim 10\%$ of the oscillation period in Fig.~\ref{fig:spin-echo length oscillation}.
Hence, over the averaging window defined by $\Delta k$, the polarization and reflectivity oscillates wildly in the range $[-1,1]$. Therefore, we can effectively approximate $p(\bk')$ as an uniform distribution over a period $(k-\pi/\xi, k+\pi/\xi)$, leading to the equation of the experimentally-observed polarization $ P^\perp_\mu(\bk)$ and reflectivity $ R^\perp (\bk)$:
\begin{eqnarray} 
\nonumber
    P^\perp_z(\bk) &=& \frac{2r_\uparrow (\bk) r_\downarrow(\bk)}{(r_\uparrow(\bk) +r_\downarrow (\bk))^2} \left(1+\cos (\phi_\uparrow(\bk) -\phi_\downarrow(\bk) ) \right), \nonumber \\  
    P^\perp_x (\bk)&=&P^\perp_y (\bk)=0, \\
    R^\perp (\bk) &=& \frac{r_\uparrow^2(\bk)+r_\downarrow^2(\bk) }{2}.   \nonumber
    \label{polarization spin echoed average}
\end{eqnarray}
Importantly, although $P^\perp_z(\bk) $'s shape is similar to $P^\|_z (\bk)$ but with a different oscillation amplitude, $R^\perp(\bk) $ and $R^\|(\bk) $ should agree exactly with each other. Hence, the difference between $R^\perp(\bk) $ and $R^\| (\bk)$, particularly at the wavevector of the intensity dip where the giant GH-shift is expected, as shown in Figs.~\ref{fig:data}(E) and (F), implies that there are other non-unitary processes involved when the sample magnetization is parallel to $\hat{\sf e}$.
Furthermore, we also note that the observed polarization has higher sensitivity to the magnetic and non-magnetic densities when the probed layer is a specially-engineered multi-layer sample, leading to giant GH-shifts. This suggests that entangled reflectometry in total reflection \cite{Quan2024} can be utilized in a complementary way to the usual polarized reflectometry method.

\end{document}